\documentclass[12pt]{article}

\usepackage[bookmarks=false]{hyperref}
\usepackage{times}
\usepackage{amsmath}
\usepackage{graphicx}
\usepackage{upgreek}
\usepackage{lineno}
\usepackage{authblk}

\newcommand{\ket}[1]{\left\vert #1 \right\rangle}

\title{Dynamical Observations of Self-Stabilising Stationary Light}

\author[1]{J.~L.~Everett}
\author[1]{G.~T.~Campbell}
\author[1]{Y.-W. Cho}
\author[1,3]{P.~Vernaz-Gris}
\author[1]{D.~B.~Higginbottom}
\author[1]{O.~Pinel}
\author[2]{N.~P.~Robins}
\author[1]{P.~K.~Lam}
\author[1]{B.~C.~Buchler}

\affil[1]{Centre for Quantum Computation and Communication Technology, Research School of Physics and Engineering, The Australian National University, Canberra, ACT 2601, Australia}
\affil[2]{Research School of Physics and Engineering, The Australian National University, Canberra, ACT 2601, Australia}
\affil[3]{Laboratoire  Kastler  Brossel,  UPMC-Sorbonne  Universites,  CNRS, ENS-PSL  Research  University, Collège  de  France,  4  Place  Jussieu,  75005  Paris,  France}

\date{}

\begin{document} 

\maketitle 

\begin{abstract}
Precise control of atom-light interactions is vital to many quantum information protocols. In particular, atomic systems can be used to slow and store light to form a quantum memory. Optical storage can be achieved via stopped light, where no optical energy remains in the atoms, or as stationary light, where some optical energy remains present during storage. In this work, we demonstrate a form of self-stabilising stationary light. From any initial state, our atom-light system evolves to a stable configuration that is devoid of coherent emission from the atoms, yet may contain bright optical excitation. This phenomenon is verified experimentally in a cloud of cold Rb87 atoms.  The spinwave in our atomic cloud is imaged from the side allowing direct comparison with theoretical predictions.
\end{abstract}

Coherent atom light interactions lie at the heart of many quantum information systems \cite{Hammerer:2010gsa,Saffman:2010ky}.  In particular, implementations of quantum repeaters will likely rely on mapping of photonic states onto atomic systems to enable storage of quantum information \cite{Lvovsky:2009fr,Sangouard:2011bp}. Deterministic quantum logic gates in optical systems may also rely on atomic state mapping to enable nonlinear photon-photon interactions \cite{Chang:2014fa,Venkataraman:2012do,Firstenberg:2013dj,Matsuoka2014,Volz:2014fx}. A fundamental issue facing any attempt to implement nonlinear cross phase modulation (XPM) is that the interaction is inherently very weak.  Techniques are therefore required to increase the interaction time or interaction strength to allow useful amounts of phase shift.  Interaction strength can be scaled up by choosing a nonlinear medium with strong optical interactions, such as Rydberg atoms \cite{Urban:2009kh,Pritchard:2010im}. A more general approach that works for any medium is to use smaller interaction volumes, since this increases the electric field per photon, and longer interaction times, which may be achieved by using an elongated nonlinear medium, slowing the group velocity of the light or using an optical cavity.  Various methods have been used to achieve one or both of these objectives, including photonic crystal fibres with \cite{Venkataraman:2012do} and without \cite{Matsuda:2009de} injected atoms, tapered optical fibres in atomic vapour \cite{Spillane:2008hd}, optical cavities \cite{Volz:2014fx,Reiserer:2014hf,Beck:2015ww} and slow light propagation \cite{Feizpour:2015bt,Chen:2006cf,Shiau:2011ko}.

Light may be slowed using Electromagnetically Induced Transparency (EIT) \cite{Fleischhauer:RevEIT:2005}.  In this scheme, a pulse of probe light resonant with an ensemble of three-level $\Lambda$-type atoms may be transmitted through the atoms with the assistance of a bright, copropagating control field that couples the probe field to a spinwave in the atomic ensemble. As the control field amplitude is reduced, the probe field velocity and amplitude are both reduced. In the limit where the control field is switched off, the probe field amplitude and velocity both fall to zero and the probe field is fully mapped into a stationary atomic spinwave. This is sometimes referred to as  \textit{stopped light}.

Another form of light with reduced group velocity is  \textit{stationary-light} (SL). This was originally proposed \cite{Andre2002} and demonstrated \cite{Bajcsy2003} by adding a counterpropagating control field to regular EIT.  The bi-directional control field leads to a stationary probe field that has non-zero amplitude. At first this effect was attributed to the standing wave in the control field creating a band-gap preventing the propagation of the probe \cite{Bajcsy2003,Andre2005}. Later analysis showed that counter-propagating control fields of very different frequencies can also give rise to SL. The explanation is that shape-preserving EIT propagation in both directions can add up to prevent propagation \cite{Moiseev2006d,Moiseev2007}. In fact, true standing wave control fields can cause unwanted coupling between counter-propagating fields and additional decay of the SL pulse  \cite{Zimmer2006,Hansen2007,Wu2010e,Peters2012a}.
The behaviour of EIT stationary light is well understood \cite{Bao2011,Hansen2007a,Moiseev2014,Wu2008,Wu2010d,Zhang2011,Zhang2012c,Zhang2013,Zhang2015}, and the interaction between SL and a stored light pulse has been demonstrated \cite{Chen2012}.

\begin{figure}[ht]{\includegraphics[width=20cm]{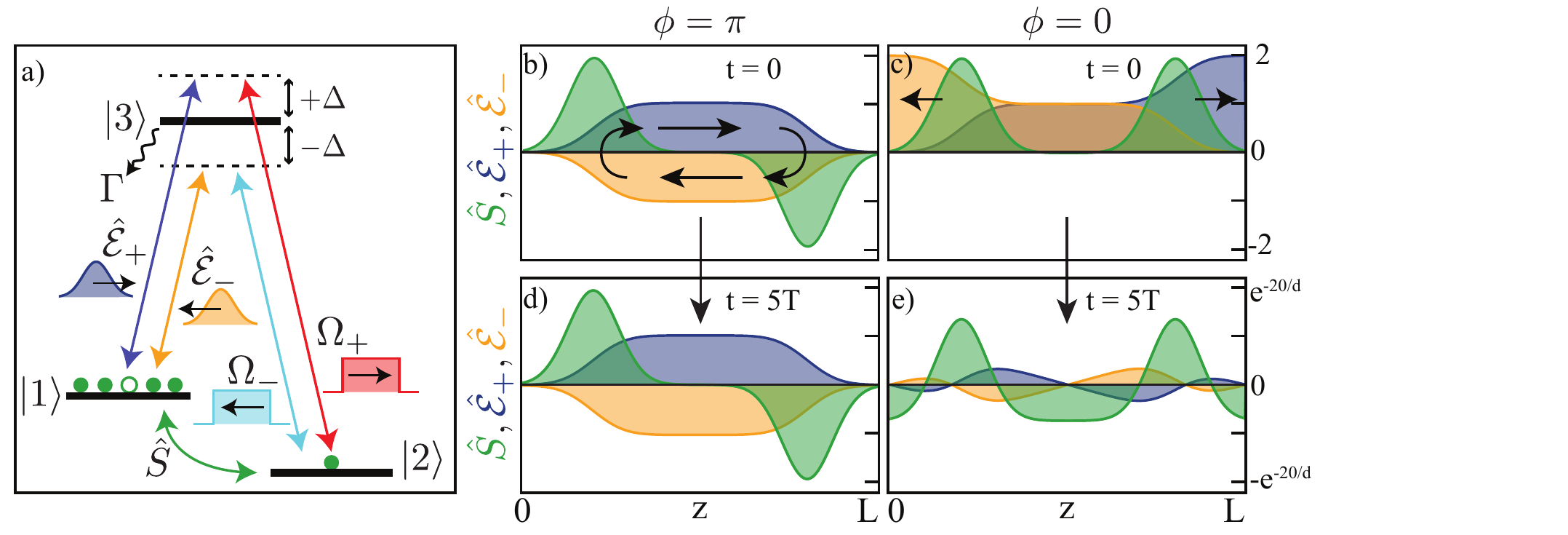}}
\caption{(a) An energy level scheme showing how the forward (+) and backward (-) probe fields are coupled with the atoms via the forward  and backward control fields. S, the spinwave, is the coherence between levels $|1\rangle$ and $|2\rangle$. Examples of (b) stationary and (c) non-stationary spinwaves at the instant the control fields are switched on. Probe fields are given by the spatial integrals of the spinwave (green) in their directions of travel, forward (blue) and backward (orange) (See Eqs.~\ref{twoleveleqns4} and \ref{twoleveleqns5}). Where the two regions of spinwave have opposite phase (b,d), the spinwave integrates to zero and does not evolve. The spinwaves at each end of the ensemble radiate probe light in both directions with opposite phase and the light emitted by one is absorbed by the other. Where the two regions have the same phase (c), they emit in phase, and the probe fields escape. As the two probe fields do not cancel, the spinwave rapidly evolves until it integrates to zero as in (e). In this example, the spinwaves were written using a probe pulse of amplitude 1 and duration T. The control field intensity is chosen so that the SL amplitude is equal to that of the input pulse.}
\label{theory}
\end{figure}

In this work we present a technique where we excite an atomic spinwave that self-stabilises to a solution that supports SL. The scheme is based on an ensemble of three-level $\Lambda$-type atoms with off-resonant driving fields (Fig.~\ref{theory}a). In this Raman configuration, bright counter-propagating control fields ($\Omega_{\pm}$) and weak probe fields ($\mathcal{E}_{\pm}$) drive the spin coherence in the atoms through a coherent scattering process. The SL states in our scheme have a significantly different nature to those observed previously using EIT. The wide class of self-stabilised states allow great flexibility in the spatial properties of the SL states.  Furthermore, unlike EIT, the spinwave and SL fields are spatially separated in the atomic medium. 

In the following section we will show theoretically how an initial spinwave encoded in the atoms evolves to a state with constant spinwave and SL amplitudes. We then present experimental results that verify the existence of this self-stabilised state by directly imaging the spin coherence as it evolves in the atomic ensemble and find excellent agreement with a simple model of the system.

\paragraph*{Theory}

The simplified atomic level scheme that we consider is illustrated in Figure \ref{theory} (a). Two hyperfine states, $\ket{1}$ and $\ket{2}$, are coupled through two Raman transitions via the the excited state $\ket{3}$. Each of the Raman transitions consists of a weak probe $\mathcal{E}_{\pm}$ on the $\ket{1} \rightarrow \ket{3}$ transition and a strong coupling field $\Omega_{\pm}$ on the $\ket{2} \rightarrow \ket{3}$ transition, where the subscripts $+$ and $-$ refer to the direction of propagation. The two Raman transitions are in two-photon resonance with the $\ket{1} \rightarrow \ket{2}$ spin coherence, counter-propagate with respect to each other, and are symmetrically detuned above and below  $\ket{3}$ by $\Delta$. The symmetric detuning ensures that the dispersion experienced by each of the probe fields due to the excited state is equal and opposite such that the two Raman transitions can be phase-matched throughout the ensemble.

The evolution of the coupled light-atom system is governed by the Maxwell-Bloch equations. We adiabatically eliminate $\ket{3}$ by assuming that $\Delta \gg \Gamma$, where $\Gamma$ is the excited state linewidth, and write the simplified equations of motion as
\begin{align}\label{twoleveleqns3}
\partial_t\hat{S}(t,z)&=i\sqrt{d}\,\Gamma\left(\frac{\Omega_+}{\Delta}\hat{\mathcal{E}}_++\frac{\Omega_-}{\Delta}\hat{\mathcal{E}}_-\right)-\gamma\hat{S}\\\label{twoleveleqns4}
\partial_z\hat{\mathcal{E}}_+(t,z)&=  i\sqrt{d}\frac{\Omega_+}{\Delta}\hat{S}\\
\partial_z\hat{\mathcal{E}}_-(t,z)&= - i\sqrt{d}\frac{\Omega_-}{\Delta}\hat{S}\label{twoleveleqns5}
\end{align}
where $d$ is the amplitude optical depth and $S(t,z)$ is the collective operator for the $\ket{1} \rightarrow \ket{2}$ spin coherence. The rate $\gamma$ is the decay of the spinwave. To understand the underlying coherent dynamics, we will ignore this decay for the time being. The derivation and assumptions necessary to obtain (\ref{twoleveleqns3}-\ref{twoleveleqns5}) are included in the supplementary materials.
Before solving these equations we note that Eqs.~\ref{twoleveleqns4} and \ref{twoleveleqns5} can be interpreted to give some intuition about the behaviour of the coupled spinwave-optical system.  They show that the amplitude of the probe field $\mathcal{E}_{-(+)}(z_0)$ is given by the spatial integral of the spinwave amplitude over the domain $(-)\infty$ to $z_0$. That means if we have an atomic ensemble over which the spinwave amplitude integrates to zero, it leads to a cancellation of the probe field amplitude at the edge of the ensemble.

We solve the equations of motion by integrating (\ref{twoleveleqns4},\ref{twoleveleqns5}), take $\Omega_+=\Omega_-=\Omega$ and substitute into (\ref{twoleveleqns3}) to arrive at
\begin{equation}
\partial_t\hat{S}(z) =-d\,\Gamma\frac{\Omega^2}{\Delta^2}\left( \int_0^z{\hat{S}(z')\mathrm{d}z'} - \int_L^z{\hat{S}(z')\mathrm{d}z'}  \right) = -d\,\Gamma\frac{\Omega^2}{\Delta^2}\int_0^L{\hat{S}(z')\mathrm{d}z'}.\label{spinwaveevolution}
\end{equation} 
The righthand side is proportional to the integral of the spinwave amplitude over the length of the ensemble, which means that the time derivative of $\hat{S}(z)$ is equal at all points in the ensemble. Consequently, the integrated spinwave amplitude will evolve at the rate $d\,\Gamma \Omega^2/\Delta^2$ towards a state where the integrated amplitude of $\hat{S}(z)$ is zero.

At one extreme we may consider a spinwave with constant amplitude over the whole ensemble. With the control fields on, this spinwave will be mapped into counterpropagating probe fields as per Eqs.~\ref{twoleveleqns4} and \ref{twoleveleqns5}.  The amplitude of the spinwave decays uniformly at rate $d\,\Gamma \Omega^2/\Delta^2$ until it reaches zero. A constant amplitude spinwave is thus a bright state of the system leading to complete emission of the stored probe light.
At the other extreme, any spinwave that is spatially orthogonal to the bright state will not evolve. In this case we a dealing with a spinwave with an integrated amplitude of zero. As discussed above, this leads to a cancellation of the probe fields at the edge of the ensemble. This cancellation means there will be no coherent emission of probe light, although inside the ensemble, the spinwave and optical fields remain constant. This is therefore a dark state of the system.

In our experiments we consider two initial spinwave configurations. In one we have equal amplitude excitations at either end of the atomic ensemble with opposite phase ($\phi=\pi$).  This is a stationary, dark state. Figure \ref{theory} (b,d) illustrates this stationary solution. The optical fields produced in each direction by each of the Gaussian spinwaves destructively interfere by the end of the ensemble so that no probe field can escape the atoms. In the second example we have Gaussian excitations with equal phase ($\phi=0$). In this case we expect that the system will evolve as illustrated in Fig.~\ref{theory} (c,e). The initial unstable spinwave self-stabilises to a dark state where there is no further coherent emission of probe light. We can understand this behaviour if we consider the initial spinwave as a sum of the bright state and an orthogonal dark state. The bright state is just a constant spinwave with an amplitude that is the mean of the spinwave.  This component decays leaving only the dark state. In the experimental results presented below we observe precisely this predicted behaviour, modified only by the incoherent decay of our atoms that we ignored in this discussion.

\paragraph*{Methods}
\begin{figure}[ht]{\includegraphics[width=0.7\columnwidth]{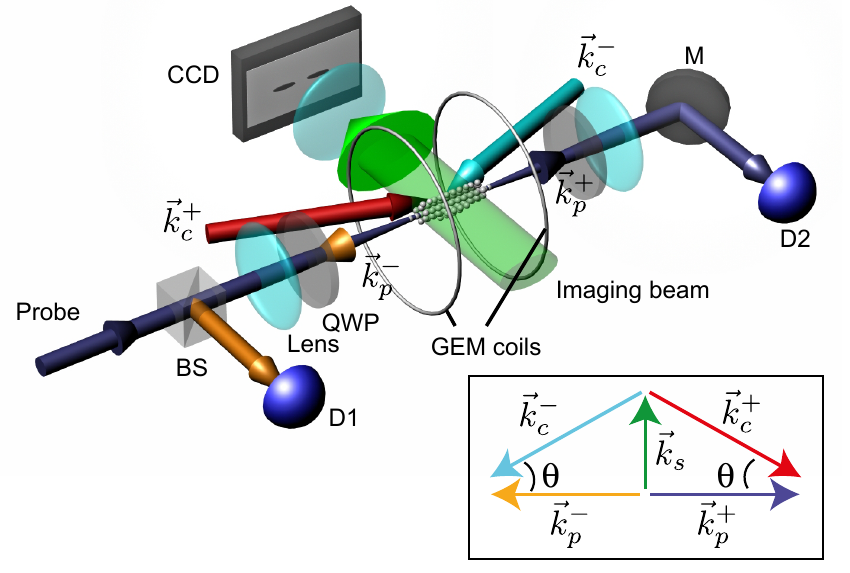}}
\centering
\caption{The experimental schematic for the SL pulse generation in an ensemble of cold $^{87}$Rb atoms. The spinwave is created by storing a probe pulse via the gradient echo memory technique with forward control field ($k_c^+$). The magnetic field gradient over the ensemble is controlled by a set of coils (GEM coils). An imaging beam (resonant with $\ket{5^2S_{1/2}(F =1)} \rightarrow \ket{5^2P_{3/2}(F' = 2)})$ is used for the absorption imaging of spin-wave.
Inset: The angle between forward (backward) probe and forward (backward) control fields are set to be $\uptheta$, where the phase matching condition $\vec{k}_p^+-\vec{k}_c^+=\vec{k}_p^--\vec{k}_c^-=\vec{k}_s$ is satisfied. Here, $\vec{k}_s$ denotes the
spinwave vector.}
\label{setup}
\end{figure}

We test the stationary solutions using a cloud of cold rubidium atoms that are confined and cooled using an elongated magneto-optical trap (MOT) as shown in Fig.~\ref{setup}a. The level scheme employed is shown in Fig.~\ref{setup}b. The atoms are optically pumped into the $\ket{1}$ state. This provides an atomic ensemble with a resonant amplitude optical depth of $\mathbf{200\pm 10}$ for the $\sigma^+$-polarised transition from $\ket{1}$ to $\ket{3}$. The three-level $\Lambda$ system is then completed by a $\sigma^-$-polarised transition from $\ket{2}$ to $\ket{3}$. The SL scheme uses counter-propagating probe fields that are symmetrically detuned by $\pm160$ MHz about the $\ket{1}\rightarrow\ket{3}$ transition and corresponding control fields that are symmetrically detuned from the $\ket{2}\rightarrow\ket{3}$ transition. A $6$~mrad phase matching angle between each probe field and its corresponding control field accounts for the 6.8 GHz frequency difference, as illustrated in Fig.~\ref{setup}c.

The two probe fields are mode-matched to the transverse profile of the atomic cloud, exactly counter-propagate and are measured with photo-detectors. The orthogonally-polarised control fields are larger in diameter to uniformly illuminate the interaction region. The bright control fields are then filtered out using a combination of spatial and polarisation filtering. Absorption imaging of the transverse profile of the atom cloud is done with an imaging pulse resonant with the $\mathbf{\ket{2}\rightarrow\ket{5^2S_{3/2}; F = 2}}$ transition on a CCD camera. This pulse illuminates the entire atom cloud from the side, as shown in Fig.~\ref{setup} (a), and allows us to image the population of atoms in the $\ket{2}$ state.

The gradient echo memory (GEM) technique \cite{Hosseini:2012go} is used to write the initial spinwaves into the $\ket{1}\rightarrow\ket{2}$ coherence. This method has the advantage that we may precisely engineer the spatial profile of the spinwave amplitude by manipulating the spectrum of the input probe light. In GEM a longitudinal magnetic field gradient is applied to the ensemble so that the two-photon detuning of the Raman transition varies linearly along the length of the atom cloud. With the forward-propagating control field present, an incoming forward-propagating probe field will be coherently scattered into a spinwave excitation at a longitudinal location that is determined by the probe frequency. The spatial profile of the stored spinwave will thus be given by the Fourier transform of the input probe pulse \cite{Buchler:2010vv,Sparkes:2012he}.

Spinwaves to test the stationary solutions are written into the atoms by storing probe pulses that contain two carrier frequencies, each with the same Gaussian envelope. By tuning the carrier frequencies of these pulses we can excite a spinwave that has a Gaussian excitation at each end of the ensemble. The relative phase, $\phi$, of these Gaussian excitations can be chosen by tuning the relative phase of the carrier frequencies in the probe pulse. Once an initial spinwave is written into the ensemble, the counter-propagating control fields are turned on and the theoretical predictions are tested by observing the output probe light and by directly imaging the evolution of the spinwave. Absorption imaging of the spinwave is similar to that done in \cite{Zhang2009}.  Further details are provided in the supplementary material.

\paragraph*{Results and Discussion}
\begin{figure}[tbp]{\includegraphics[width=0.90\columnwidth]{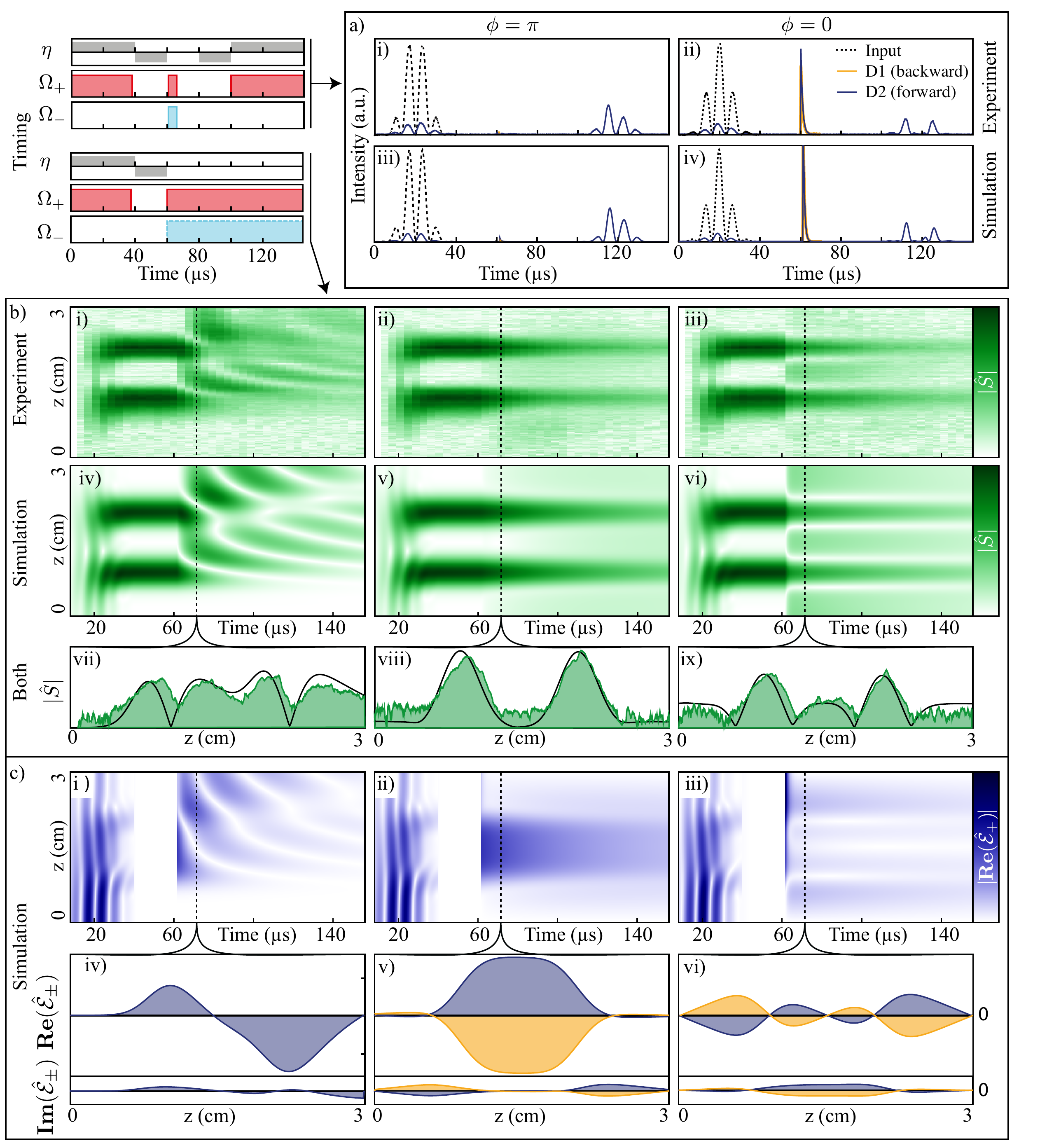}}
\centering
\caption{Measurement and simulation results showing: (a) the forward- and backward-propagating probe fields and (b) the atomic coherence spinwave for SL experiments. The simulated optical field for the same conditions as (b) are shown in (c). In (b) and (c) the left column shows the evolution with only the forward control field. The middle column shows the evolution of a spinwave that is initialised to be stationary ($\phi=\pi$). The right column shows the evolution of a spinwave that is not initially stationary ($\phi=0$). The inset (top-left) shows timing diagrams for the experiments (a) and (b). For the simulations the control field Rabi frequency was $\Omega_\pm = 2\pi\times (2.4\pm 0.1)$~MHz and the optical depth and gradient parameters are measured from the experiment. The ground-state decay $\gamma_0$=500~Hz was based on previous experiments in the same MOT \cite{Cho:16}. A small difference in the simulation rephasing time is introduced to match the experimental rephasing.
}
\label{results}
\end{figure}

We will now compare our experimental results to the simple analytic theory presented above and numerical simulations that include decoherence effects (see supplementary material). Figure~\ref{results}(a) shows the photo-detector measurements of the light that is released by the atoms (D1 and D2 in Fig.~\ref{setup}). Upon input, a small amount of leakage is observed at D2 due to the finite optical depth. The ensemble is then illuminated by the dual control fields at 60~$\mu s$ and for the $\phi=0$ case we see a large signal at D1 and D2.  We can attribute this signal to the evolution of the non-stationary spinwave as illustrated in Fig.~\ref{theory}(c,d).  Upon recall from the ensemble with the forward control beam, we observe a much smaller output for the  $\phi=0$ case.  The numerical simulation is in very good agreement with these observations.

A more detailed picture of the dynamics is gained from the absorption imaging shown in Fig.~\ref{results}(b). The experimental data in (i-iii) is gathered by repeatedly running the experiment and moving the timing of the imaging beam to build up a picture of the spinwave evolution. The results of numerical simulations that solve the equations of motion with decay for the three-level system are shown directly underneath each experimental image. Figure~\ref{results}(c) shows the simulated optical field envelope within the ensemble, illustrating the SL aspect of the solution.  

Our experimental data shows excellent agreement with the behaviour predicted by the simulations. The middle column of Fig.~\ref{results} shows results and simulations for a stationary spinwave, where the integrated amplitude of the initial spinwave is zero. In this case, as expected, we observe no evolution of the shape of the spinwave. The right-hand column, where the initial spinwave is not stable, shows a rapid evolution to a stationary spinwave in both the experimental data and simulated results. Once they have evolved to a steady-state, the spinwaves show a strong resemblance to those that we predict from the simple model, as can be seen by comparing the illustrations in Fig.~\ref{theory} to the measured and simulated spinwaves in Fig.~\ref{results} (b; viii,ix).  Once the spinwave has reached a steady-state in terms of shape, it continues to decay at a rate given by the control field scattering. The spinwave decay rate roughly agrees between experiment and simulation at $10\pm1~$kHz measured and $12~$kHz in the simulations.

We also verify that the steady-state solutions are a result of interference between the counter-propagating Raman transitions by repeating the experiment without turning on the backward control field. In this case, shown in the left hand column of Fig.~\ref{results}(b,c), the free-induction decay oscillation can be seen as the probe propagates in a medium with a narrow-linewidth absorption feature.

While the results closely match the simulations, there are slight discrepancies from the simplified model of the system. In the experiment the probe fields undergo incoherent absorption by the atoms while travelling across the memory and are no longer equal at every position. This causes slight changes in the shape of the spinwave. This is particularly evident in Fig \ref{results} (b; ii, v), where the edges of the spinwave gradually increase in amplitude. In the limit where $\frac{d\Gamma}{\Delta}\ll 1$, the mechanism causes an additional decay of the SL field with exponent $-\left(\frac{d\,\Gamma \,\Omega}{\Delta^2}\right)^2$. In the simulation this evolution is accompanied by a small amount of probe leakage. The leakage was below the experimental background noise level and could not be measured. A further discrepancy is that the predicted rate of evolution of the bright state is 0.8~MHz but is measured to be $0.6\pm0.1$~MHz based on the detected probe field, which we attribute to a slight inaccuracy of the phase-matching angle.

At larger single-photon detuning $\Delta$, the only important decay mechanisms are the spinwave decoherence, caused by atomic motion and ground-state dephasing, and control field scattering. The equations of motion for this system approach the simplified eqs \ref{twoleveleqns3}- \ref{twoleveleqns5}, with a decay term $\gamma=\gamma_0 + 2\Gamma\Omega^2/\Delta^2$.

It is interesting to consider how our SL scheme may perform in an implementation of XPM, where one optical field creates an AC-Stark shift of an atomic level with which another optical field is interacting \cite{Schmidt:1996wh}. As discussed in the introduction, this capability is highly appealing for quantum information processing but the effect is typically too weak to be useful.  A reasonable benchmark for the improvement of XPM due to slow or stationary light effects is the magnification of the XPM phase shift compared to that caused by a single photon propagating freely through the ensemble.

In our system the spatial configuration of the SL within the ensemble brings some advantages. The light that is to be phase-shifted may be stored as a spinwave at the centre of the ensemble. SL can then be generated using the configuration of Fig.~\ref{theory}b to overlap the SL with the stored spinwave. The frequencies are chosen so that one or both of the SL optical fields creates an AC-Stark shift of the stored spinwave. 

The enhancement factor arising from SL comes from an increased time-intensity product. Both SL intensity and decay rate depend on control field intensity, while SL intensity also scales linearly with the optical depth, $d$. Figs.~\ref{theory}b,d illustrate an initial intensity of the SL equal to the average intensity of the input photon, and the subsequent decay (see supplementary materials for a calculation of the enhancement factor). The intensity of a single photon is also governed by the geometry, so we require the interaction volume to be as small as possible. To maximise XPM therefore requires careful experimental design to minimise the beam volumes and maximise the density of the atomic cloud

Scaling of XPM with $d$ is common to EIT schemes \cite{Lukin2000,Wang2006,Feizpour2016}, where both pulses travel through the atomic ensemble. These schemes can require careful matching of the pulse velocities to maximise the interaction time. There is also an argument that the locality of the AC-Stark shift effect causes phase noise and limits the total phase shift in some multimode systems \cite{Shapiro2006,Gea-Banacloche2010}. This is not a problem for single mode systems, such as a cavity, where the whole cavity eigenmode is uniformly shifted \cite{Hickman:2015dq}. In our proposed XPM scheme neither pulse is moving, and matching group velocities is not an issue. Furthermore, the SL cycles between the spinwaves at either end of the atomic ensemble leading to uniform interaction with the stored spinwave just as if we were using a cavity.

A recent experiment demonstrated an enhancement of $d$ in an EIT scheme, where a probe photon modulated a level to which another weak pulse coupled \cite{Feizpour:2015bt}. Phase shifts of $13\pm1\upmu$rad per photon were observed even though the geometry of that experiment limited the available optical depth. It is possible to achieve similar optical intensities in our cold-atom cloud as in the above mentioned experiment, while making use of our full optical depth. We predict that phase shifts in the order of milliradians are possible.

\paragraph{Conclusion}
We have demonstrated a novel form of stationary light and a new type of behavior for three-level atomic systems. The stationary case is promising to explore for producing non-linearities, while the non-stationary case also produces interesting and novel dynamics.

 \bibliography{Bibliographies-Stationarylightbib}
 \bibliographystyle{utphys}

\pagebreak

\section*{Dynamical Observations of Self-Stabilising Stationary Light: Supplementary Materials}
\section{Interaction of three-level atoms with counter-propagating fields}
We follow the derivation in \cite{Gorshkov2007c} and add counter-propagating fields with the assumption that each probe field interacts only with the control field that co-propagates with it. The operators for two weak counter-propagating quantum fields and two counter-propagating classical control fields are
\begin{eqnarray}
\hat{\textbf{E}}_{p+}(z)&=&\epsilon_{p+}\left(\frac{\hbar\omega_{p+}}{4\pi c \epsilon_0A}\right)^{1/2}\int_{\omega_{p+}} \mathrm{d}\omega\left(\hat{a}_\omega e^{i\omega z/c}+\hat{a}^\dag_\omega e^{-i\omega z/c}\right)\nonumber\\
\hat{\textbf{E}}_{p-}(z)&=&\epsilon_{p-}\left(\frac{\hbar\omega_{p-}}{4\pi c \epsilon_0A}\right)^{1/2}\int_{\omega_{p-}} \mathrm{d}\omega\left(\hat{a}_\omega e^{-i\omega z/c}+\hat{a}^\dag_\omega e^{i\omega z/c}\right)\nonumber\\
\textbf{E}_{c+}(z)&=&\epsilon_{c+}\mathcal{E}_{c+}(t-z/c)\textrm{Cos}[\omega_{c+}(t-z/c)]\nonumber\\
\textbf{E}_{c-}(z)&=&\epsilon_{c-}\mathcal{E}_{c-}(t+z/c)\textrm{Cos}[\omega_{c-}(t+z/c)]\nonumber
\end{eqnarray}
where we assume optical modes that each exist in a small bandwidth around a carrier frequency given by $\omega_{p+}=\omega_{13}+\Delta^++\delta$, $\omega_{p-}=\omega_{13}+\Delta^-+\delta$, $\omega_{c+}=\omega_{23}+\Delta^+$, $\omega_{c-}=\omega_{23}+\Delta^-$. The interaction part of the Hamiltonian is then

\begin{align}
\hat{V}=-\hbar\sum^{N}_{i=1}\Bigg[\left(\Omega_{c+}(t-z_i/c)e^{-i\omega_{c+}(t-z_i/c)} + \Omega_{c-}(t+z_i/c)e^{-i\omega_{c-}(t+z_i/c)}\right)\hat{\sigma}^i_{32}  \nonumber\\ 
+g\left(\frac{L}{2\pi c}\right)^{1/2}\left(\int_{\omega_{p+}} \mathrm{d}\omega\hat{a}_\omega e^{i\omega z/c}\hat{\sigma}^i_{31} + \int_{\omega_{p-}}\mathrm{d}\omega\hat{a}_\omega e^{-i\omega z/c}\hat{\sigma}^i_{31}\right)+ \textrm{H.c.}\Bigg].
\end{align}
We define slowly varying collective operators

\begin{align} 
\hat{\sigma}_{\mu\mu}(z,t)&= \frac{1}{N_z}\sum^{N_z}_i\hat{\sigma}^i_{\mu\mu}(t) \nonumber\\
\hat{\sigma}_{32}^\pm(z,t)&=\frac{1}{N_z}\sum^{N_z}_i\hat{\sigma}_{32}^i(t)e^{-i\omega_{c\pm}(t\mp z_i/c)} \nonumber\\
\hat{\sigma}_{31}^\pm(z,t)&=\frac{1}{N_z}\sum^{N_z}_i\hat{\sigma}_{31}^i(t)e^{-i\omega_{p\pm}(t\mp z_i/c)} \nonumber\\
\hat{\sigma}_{21}^\pm(z,t)&=\frac{1}{N_z}\sum^{N_z}_i\hat{\sigma}_{12}^i(t)e^{-i(\omega_{p\pm}-\omega_{c\pm})(t\mp z_i/c)} \nonumber\\
\hat{\mathcal{E}}_\pm(z,t)&=\sqrt{\frac{L}{2\pi c}}e^{i\omega_{p\pm}(t\mp z/c)}\int_{\omega_{p\pm}} \mathrm{d}\omega\hat{a}_\omega(t) e^{\pm i\omega z/c} \nonumber \\
\end{align}
by assuming that each slice $dz$ of the ensemble contains a number of atoms $N_z \gg 1$. The commutators for the collective operators are

\begin{align}
\left[\hat{\sigma}_{\mu\nu}(t),\hat{\sigma}_{\alpha\beta}(t)\right]&=\delta_{\nu\alpha}\hat{\sigma}_{\mu\beta}(t) - \delta_{\mu\beta}\hat{\sigma}_{\alpha\nu}(t)\nonumber\\
\left[\hat{\mathcal{E}}_\pm(t),\hat{\mathcal{E}}_\pm^\dag(t)\right]&=1. \nonumber
\end{align}
The slowly varying operators are inserted into $\hat{V}$, giving a Hamiltonian

\begin{eqnarray}
\hat{H}= \int d\omega\hbar\omega\hat{a}_\omega^\dag\hat{a}_\omega - \hbar\omega_{p+}\frac{1}{L}\int_0^L dz\hat{\mathcal{E}}_+^\dag\hat{\mathcal{E}}_+-\hbar\omega_{p-}\frac{1}{L}\int_0^L dz\hat{\mathcal{E}}_-^\dag\hat{\mathcal{E}}_-+ \nonumber\\
\int^L_0dz\hbar n(z) \times \Bigg(\Delta_+\hat{\sigma}_{33}^++\Delta_-\hat{\sigma}_{33}^--\Bigg[\Omega_{c+}(t-z/c)\hat\sigma^+_{32} + \Omega_{c-}(t+z/c)\hat{\sigma}^-_{32} + \nonumber\\\Omega_{c+}(t-z/c)\hat\sigma^-_{32}e^{-i[\omega_{c+}(t-z/c)-\omega_{c-}(t+z/c)]} + \Omega_{c-}(t+z/c)\hat{\sigma}^+_{32}e^{-i[\omega_{c-}(t+z/c)-\omega_{c+}(t-z/c)]} \nonumber\\
+g\Big(\hat{\mathcal{E}}_+\hat{\sigma}^+_{31}+\hat{\mathcal{E}}_-\hat{\sigma}^-_{31} + \hat{\mathcal{E}}_+\hat{\sigma}^-_{31}e^{-i[\omega_{p+}(t-z/c)-\omega_{p-}(t+z/c)]}\nonumber\\
+\hat{\mathcal{E}}_-\hat{\sigma}^+_{31}e^{-i[\omega_{p-}(t+z/c)-\omega_{p+}(t-z/c)]}\Big)+ \textrm{H.c.}\Bigg]\Bigg),
\end{eqnarray}
where the $(z,t)$ dependence of the operators is omitted for readability.

We ignore cross-terms of the form $\Omega_{c\pm}\hat{\sigma}^\mp_{32}$ and $\hat{\mathcal{E}}_\pm\hat{\sigma}^\mp_{31}$. This discards a rapid variation in the AC-Stark shift due to the beat-note between the two control fields \cite{Johanreference} and also the possibility of creating additional coherences of higher spatial frequencies as described in \cite{Wu2010e}. For $\Delta$ much greater than the probe bandwidth, the rapid variation in the AC-Stark shift will average to zero.

The excited state is written as two operators $\hat{\sigma}^+_{33}$ and $\hat{\sigma}^-_{33}$, to account for the separate rotating frames used for the two probe fields.

As $\omega_{p+} - \omega_{c+} = \omega_{p-}-\omega_{c-}$, the two spinwave operators $\hat{\sigma}^\pm_{12}$ have spatial dependence given by $\vec{k}_{p\pm}-\vec{k}_{c\pm}$. For $|\vec{k}_{c\pm}| > |\vec{k}_{p\pm}|$, it is possible to set $\vec{k}_{p+}-\vec{k}_{c+}\approx\vec{k}_{p-}-\vec{k}_{c-}$ and $\vec{k}_{p+} \parallel \vec{k}_{p-}$ using angle phase-matching. These two quantities cannot be matched exactly as the two control fields have different frequencies. However, for the experimental parameter of $\Delta=160$~MHz, the $k$-vector mismatch due to the difference in detunings is negligible: $\vec{k}_{s+}-\vec{k}_{s-}\approx 10^{-6}\vec{k}_{s+}$.

Finally, we arrive at the standard three-level equations, but with an extra pair of counter-propagating fields and optical coherences.
\begin{eqnarray}
\partial_t\hat{\sigma}_{13}^+&=& -(\Gamma+ i\Delta_+)\hat{\sigma}^+_{13}+i g \hat{\mathcal{E}}_+ + i\Omega_{c+}\hat{\sigma}_{12} \label{eq:appdx:1}\\
\partial_t\hat{\sigma}_{13}^-&=& -(\Gamma+ i\Delta_-)\hat{\sigma}^-_{13}+i g \hat{\mathcal{E}}_- + i\Omega_{c-}\hat{\sigma}_{12} \label{eq:appdx:2}\\
\partial_t\hat{\sigma}_{12}&=& -(\gamma+i\delta)\hat{\sigma}_{12}  +i\Omega_{c+}^*\hat{\sigma}^+_{13}+i\Omega_{c-}^*\hat{\sigma}^-_{13} \label{eq:appdx:3}\\
(\partial_t+ c\partial_z)\hat{\mathcal{E}}_+ &=& igN\hat{\sigma}^+_{13} \label{eq:appdx:4}\\
(\partial_t- c\partial_z)\hat{\mathcal{E}}_- &=& igN\hat{\sigma}^-_{13} \label{eq:appdx:5}
\end{eqnarray}

\section{Approximations and simplified equations}

Solutions to equations (\ref{eq:appdx:1}-\ref{eq:appdx:5}) can be found more easily by eliminating the time derivatives in eqs.~(\ref{eq:appdx:4},\ref{eq:appdx:5}). This corresponds to assuming infinite propagation velocity for light in the absence of atoms in the ensemble. For our experiment, the propagation time $L/c$ is on the order of $10^{-10}$~s, far shorter than any other timescale in the system, and the time-derivatives in eqs.~(\ref{eq:appdx:4},\ref{eq:appdx:5}) can be neglected.

The equations can be further simplified, and written in terms of parameters that are directly measurable from experiment, by renormalising the collective operators by defining $\hat{S}=\sqrt{N}\hat{\sigma}_{12}$ and  $\hat{P_\pm}=\sqrt{N}\hat{\sigma}^\pm_{13}$. The equations can then be written in terms of the optical depth $d$ and a normalised length $\xi$ which runs from $0$ to $1$, giving
\begin{eqnarray}
\partial_t\hat{P}_\pm&=& -(\Gamma+ i\Delta_\pm)\hat{P}_\pm+i \sqrt{d}\Gamma \hat{\mathcal{E}}_\pm + i\Omega_{c\pm}\hat{S} \nonumber\\
\partial_t\hat{S}&=& -(\gamma+i\delta)\hat{S}  +i\Omega_{c+}^*\hat{P}_++i\Omega_{c-}^*\hat{P}_- \nonumber\\
\partial_\xi\hat{\mathcal{E}}_\pm &=& \pm i\sqrt{d}\hat{P}_\pm\nonumber.
\end{eqnarray}

The excited state may be adiabatically eliminated by assuming that its evolution is dominated by the detuning $\Delta\gg\partial_t\hat{P}_\pm$ so that
\begin{eqnarray}
\hat{P}_\pm=i\left(\sqrt{d}\Gamma\hat{\mathcal{E}}_\pm+\Omega_{c\pm}\hat{S}\right)/\left(\Gamma+i\Delta\right). \nonumber
\end{eqnarray}
The equations of motion can now be expressed as
\begin{eqnarray}
\partial_t\hat{S}&=&-(\gamma'+i\delta')\hat{S}+i\sqrt{d}\Gamma\left(\frac{\Omega^*_{c+}}{\tilde{\Delta}_+}\hat{\mathcal{E}}_++\frac{\Omega^*_{c-}}{\tilde{\Delta}_-}\hat{\mathcal{E}}_-\right)\label{twoleveleqns1}\\
\partial_\xi\hat{\mathcal{E}}_\pm&=&\pm i\left(d\frac{\Gamma}{\tilde{\Delta}_\pm}\hat{\mathcal{E}}_\pm+\sqrt{d}\frac{\Omega_{c\pm}}{\tilde{\Delta}_\pm}\hat{S}\right)
\label{twoleveleqns2}
\end{eqnarray}
where
\begin{equation}
\delta'=\delta-\frac{|\Omega_{c+}|^2\Delta_+}{\Gamma^2+\Delta_+^2}-\frac{|\Omega_{c-}|^2\Delta_-}{\Gamma^2+\Delta_-^2},\ \gamma'=\gamma+\Gamma\frac{|\Omega_{c+}|^2}{\Gamma^2+\Delta_+^2}+\Gamma\frac{|\Omega_{c-}|^2}{\Gamma^2+\Delta_-^2},\ \tilde{\Delta}_\pm = \frac{\Delta_{\pm}^2+\Gamma^2}{\Delta_\pm+i\Gamma}\nonumber.
\end{equation}

Equations \ref{twoleveleqns1},\ref{twoleveleqns2} are integrated for the numerical simulations while equations (\ref{twoleveleqns3},\ref{twoleveleqns4}) are obtained by discarding decay and atomic detuning terms, setting $\Omega_{c+}=\Omega_{c-}=\Omega$, and by assuming $\Delta \gg \Gamma$, so that $\tilde{\Delta}_+\approx-\tilde{\Delta}_- \approx \Delta$. This approximation neglects off-resonant absorption, which in our experiment does lead to a slight departure from a stationary state. Numerical simulations are carried out using XMDS \cite{Dennis2013}.

The first term in equation \ref{twoleveleqns2} is a dispersion term. For $\tilde{\Delta}_+=-\tilde{\Delta}_-$, the term is common to both fields and may be included in a global phase to remove it from the equations of motion. The cancellation of the dispersion by using oppositely detuned transitions produces a time-reversal symmetry in the equations of motion. This is detailed by Moiseev and Tittel \cite{Moiseev2011}, and was useful in interpreting the dynamics of our system. 

\section{Absorption Imaging}
\begin{figure}[ht]{\fbox{\includegraphics[width=15cm]{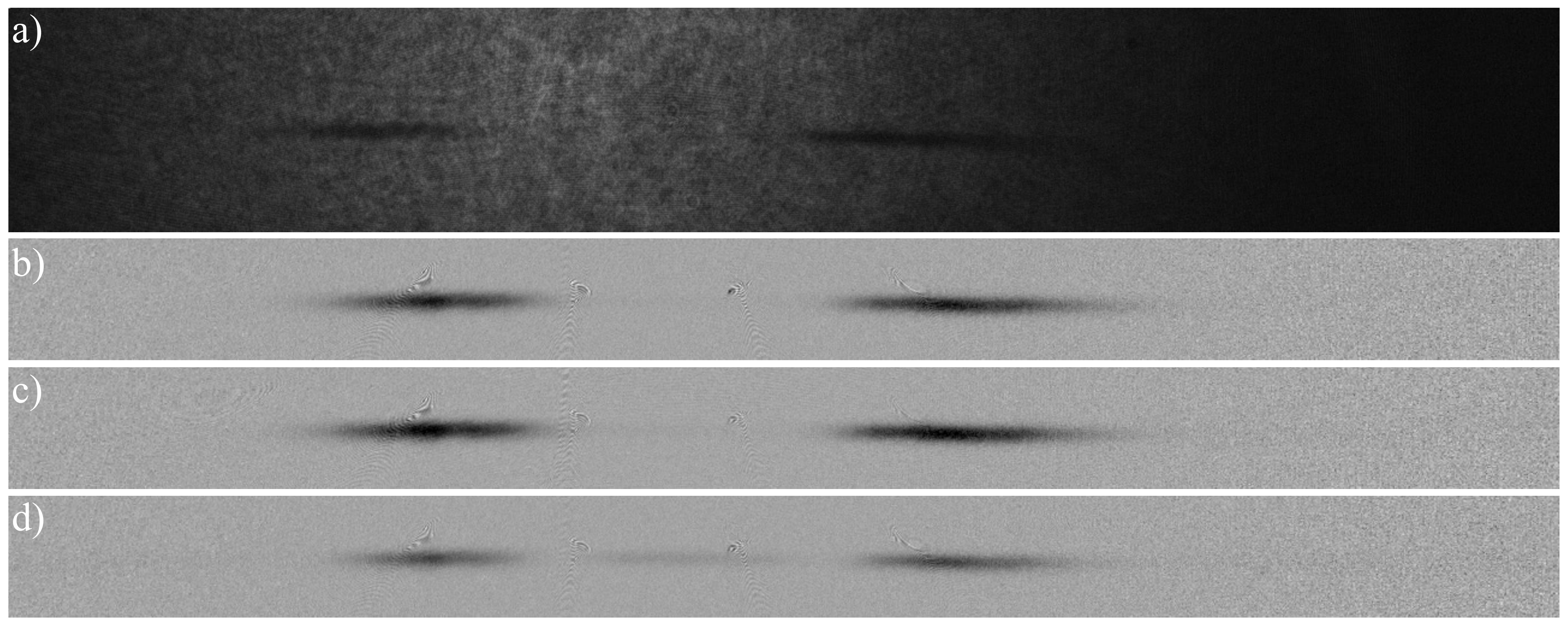}}}
\caption{(a) Raw absorption image at 58 $\upmu$s. Processed images at (b) 58$\upmu$s, (c) $\phi=\uppi$ at 62 $\upmu$s and (d) $\phi=0$ at 62$\upmu$s.}\label{fig:Imaging}
\end{figure}

We use a large (3") aperture lens to image the MOT onto a CCD camera. This is done using fluorescence from the atoms during the trapping phase of the experiment cycle. We then shine a collimated beam resonant with the $\ket{2}\rightarrow\ket{5^2P_{3/2}(F' = 2)}$ transition through the atoms at an angle that is perpendicular to the propagation axis of the probe fields. The beam illuminates the CCD camera with a shadow cast by absorption in the atomic ensemble. We can infer the relative magnitude of the spinwave from the optical depth of its shadow
\begin{equation}
  \vert\hat{S}(x,y)\vert  \propto \sqrt{\ln{[I_0(x,y)/I(x,y)]}}
\end{equation}

Spatial measurements of $\vert\hat{S}\vert$ are captured by illuminating the ensemble with a 4~$\mu$s imaging pulse. Each measurement is constructed from ten images that are averaged to reduce shot-to-shot noise of the CCD camera. Figure \ref{fig:Imaging} shows a single image capture (a) from a SL experiment along with the inferred spinwaves after the optical pulse is initially encoded in the atoms (b) and after it has evolved into a steady state (c,d). The evolution of the spinwave is reconstructed by running the experiment repeatedly and varying the timing of the imaging light.

\section{Cross-Phase Modulation}

\begin{figure}[ht]{\includegraphics[width=\linewidth]{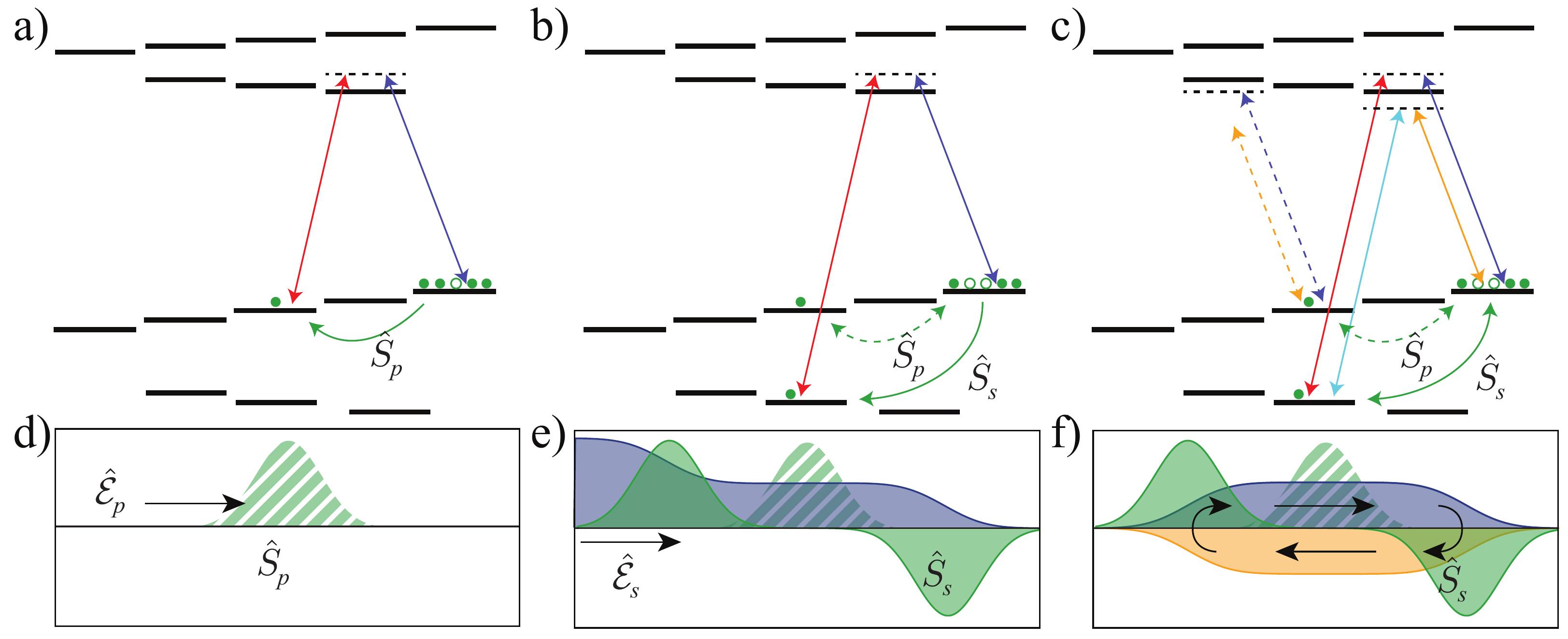}}
\caption{An implementation of cross-phase modulation using stationary light. The atomic level structure for the scheme is shown in the top row (a-c) and the spatial structure of the spinwaves and optical fields is illustrated in the bottom row (d-f). A target state is initially stored in the centre of the ensemble in the coherence between two Zeeman sub-levels using the GEM technique (a,d). A signal photon is then stored in the coherence between two hyperfine levels with a spatial mode that will lead to stationary light (b,e). Two counter-propagating control fields are used to create a stationary light optical field from the signal photon that causes an ac-Stark shift for the Zeeman level in which the target photon is stored (c,f). The signal photon and target photon are then retrieved from the ensemble.}\label{fig:XPM}
\end{figure}

We propose a scheme to use SL to implement a cross-phase modulation between two photons. The idea, illustrated in fig.~\ref{fig:XPM}, is to store a probe photon in the centre of the ensemble and then create SL in that region from a second signal photon. The probe photon would be stored in the $\ket{F=1, m_F=2} \Rightarrow \ket{F=1, m_F=0}$ coherence using the GEM technique (a,d). A modulated signal photon would then be stored on the $\ket{F=1, m_F=2} \Rightarrow \ket{F=0, m_F=0}$ coherence, as we have demonstrated here (b,e). Applying counter propagating control fields generates SL from the signal photon (c,f), resulting in an AC-Stark shift of the $\ket{F=1, m_F=0}$ state through off-resonant interaction of the signal field with the $\ket{F'=1, m_F=-1}$ state. A magnetic field would be used such that the AC-Stark shift is optimised. Both the probe field and signal field could then be retrieved from the memory (not shown). It is worth noting that the probe and signal photons could be stored and retrieved simultaneously because they are coupled to the ensemble using independent control fields or at different times.

The total cross-phase modulation achievable in the scheme can be calculated by integrating the AC-Stark shift over time as the SL decays. We assume that the only significant decay of the SL field is due to control-field scattering. The AC-Stark shift depends on the Rabi frequency and the detuning of the probe photon from the interacting atomic level:
\begin{equation}
\Delta_{AC}=-\frac{\Omega_s^2}{4\Delta_s}
\end{equation}
And, provided that $\int^{\infty}_{-\infty}\left|\hat{\mathcal{E}}(t)\right|^2 \mathrm{d}t=1$ for a single photon,
\begin{equation}
\Omega_s^2=\frac{\Gamma\sigma_{at}}{A}\left|\hat{\mathcal{E}}(t)\right|^2
\end{equation}
where $\sigma_{at}$ is the effective interaction cross-section for the Stark-shift transition, and $A$ is the cross-sectional area of the signal photon.

A probe photon of the form $\hat{\mathcal{E}}_s(t)=\frac{1}{\sqrt{2\tau}}e^{-\frac{t^2}{4\tau}}(e^{i\omega_+t} - e^{i\omega_-t})$ is stored in the memory over the time $-t_0$ to $t_0$. The sidebands $\omega_\pm$ are chosen such that the two frequencies are stored in separate halves of the memory and will generate SL. Assuming ideal GEM storage \cite{Hush2013}, we can approximate the spinwave as the Fourier transform

\begin{equation}
\hat{S}(z,t_0)=\frac{\eta\sqrt{\Gamma}}{\sqrt{2\pi}}\int_{-\infty}^{\infty}{\exp(iz\eta t)\hat{\mathcal{E}}_s(0,t)}\mathrm{d}t
\end{equation}
The transform is area preserving apart from a factor of $\sqrt{\Gamma}$:

\begin{align}
\int_0^1|\hat{S}(z,t_0)|^2\mathrm{d}z=\Gamma\int|\hat{\mathcal{E}}_s(0,t)|^2\mathrm{d}t
\end{align}
Because the two frequencies are separated in the memory, integrating over half the memory gives

\begin{align}
\int_0^{1/2}|\hat{S}(z,t_0)|^2\mathrm{d}z=\frac{\Gamma}{2}\int|\hat{\mathcal{E}}_s(0,t)|^2\mathrm{d}t
\label{spinwavearea}
\end{align}
Using eqs.~(\ref{twoleveleqns3},\ref{twoleveleqns4},\ref{spinwavearea}) and integrating over half the memory to solve for $\Omega^2_s$ at the centre,

\begin{align}
\Omega_s^2 = &\frac{\Gamma\sigma_{at}}{A}\int_0^{1/2}{\left|\sqrt{d} \frac{\Omega_c}{\Delta_c} \hat{S}(z,t_0)\right|^2\mathrm{d}z\times \exp(2\gamma(t_0-t))} \\= &\frac{d\sigma_{at}\Gamma^2}{2A}\frac{\Omega_c^2}{\Delta_c^2}\exp(4\Gamma\frac{\Omega_c^2}{\Delta_c^2}(t_0-t))
\end{align}
The SL intensity and decay rates are both proportional to control field intensity, so this factor cancels in the time integral and the total cross-phase modulation generated in the scheme is
\begin{equation}
\phi_{s} = \int_{t_0}^{\infty}{\Delta_{AC}\mathrm{d}t} = -\frac{\Gamma}{ \Delta_s}\frac{\sigma_{at}}{A}\frac{d}{32}
\end{equation}

For the level scheme proposed, taking account of relative transition strengths, with the magneto-optic trap used in this experiment, and a waist of 13 $\upmu$m for the signal photon, a phase shift of 1 mrad may be achieved.

\end{document}